\documentclass[conference]{IEEEtran}
\IEEEoverridecommandlockouts
\usepackage{cite}
\usepackage{amsmath,amssymb,amsfonts}
\usepackage{algorithmic}
\usepackage{algorithm}
\usepackage{graphicx}
\usepackage{textcomp}
\usepackage{xcolor}
\usepackage{tabularx}
\usepackage[normalem]{ulem}
\usepackage{multirow}
\usepackage{needspace}
\usepackage{caption}
\usepackage{mdframed}
\usepackage{ragged2e}
\usepackage{listings}
\usepackage{subcaption}
\def\BibTeX{{\rm B\kern-.05em{\sc i\kern-.025em b}\kern-.08em
    T\kern-.1667em\lower.7ex\hbox{E}\kern-.125emX}}
\begin{document}
\bstctlcite{IEEEexample:BSTcontrol}

\title{RAG-driven Multi-Agent LLM Framework with Task Decomposition for Beyond 5G Auto-Configuration
}

\author{\IEEEauthorblockN{ İrşat Emin Sarıdaş\IEEEauthorrefmark{1}\IEEEauthorrefmark{2}, Onur Salan\IEEEauthorrefmark{1}\IEEEauthorrefmark{2}, Ali Görçin\IEEEauthorrefmark{1}\IEEEauthorrefmark{2}, İbrahim Hökelek\IEEEauthorrefmark{1}, Hakan Ali Çırpan\IEEEauthorrefmark{2}}

\IEEEauthorblockA{\IEEEauthorrefmark{1}  {Communications and Signal Processing Research (HİSAR) Lab., T{\"{U}}B{\.{I}}TAK B{\.{I}}LGEM, Kocaeli, Turkey}}

\IEEEauthorblockA{\IEEEauthorrefmark{2} Department of Electronics and Communication Engineering, Istanbul Technical University, {\.{I}}stanbul, Turkey}}

\maketitle

\begin{abstract}
While Large Language Models (LLMs) offer a promising path toward intent-driven network management by translating natural language human intents into machine-readable configurations, they often suffer from hallucinations and structural inconsistencies in multi-step and complex tasks. To address these challenges, this paper proposes a retrieval-augmented and task decomposition-based multi-agent LLM framework for Beyond 5G network auto-configuration. The framework employs a semantic retrieval-augmented generation pipeline to ensure that its outputs are aligned with technical standards and vendor-specific manuals. Furthermore, it introduces a modular architecture for configuration generation, closed-loop configuration verification, and network deployment, in which complex tasks are decomposed into smaller sub-tasks handled by specialized agents. In this architecture, hallucinated configuration parameters are identified by the configuration verifier agent and corrected through low computational segment-level regeneration. The performance evaluation experiments with the OpenAirInterface emulator demonstrate that the proposed task decomposition-based configuration and verification approach improves the average success rate by 22.7\% over monolithic methods, achieving 94.4\% success in network configuration. 
\end{abstract}


\section{Introduction}

Future International Mobile Telecommunications (IMT)-2030 systems are anticipated to support extremely diverse usage scenarios by incorporating artificial intelligence (AI)-enabled capabilities across the network. In particular, International Telecommunication Union (ITU) envisions AI-native radio networks that enable automated and intelligent networking services, as well as digital twin networks for efficient verification, simulation, deployment, and management. Meanwhile, beyond 5G (B5G) network environments are becoming increasingly heterogeneous, making network configuration and operational management extremely complex and reliant on expert knowledge \cite{ITU2026IMT2030, llmfortelecom}. The evolution of network management paradigms reflects a broader effort to reduce human involvement in increasingly complex network operations. Autonomic management aims to remove the human administrator from the control loop by allowing the network to manage itself \cite{anm2012}, while recent intelligent management frameworks extend this capability by enabling networks to reason about their operational state and adapt configurations under dynamically changing conditions \cite{etsi2019zero}. 

Large language models (LLMs) have emerged as promising enablers for intent-driven network planning and management workflows, by translating natural language intents into machine-interpretable network representations \cite{llmnetworking, LLMNetConf, mekrache2024intent, coronado2022zero}. Recent studies have been exploring agentic and intent-driven AI architectures for autonomous B5G network control. The authors in \cite{agentran} propose LLM-based agents which interpret natural-language intents and decompose them across multiple timescales, spatial domains, and protocol layers to enable coordinated RAN control. Concurrently, cloud-native frameworks employ LLMs to generate configurations that are tailored to underlying scenarios while ensuring compatibility with multi-vendor open RAN deployments, guided by high-level operator intents in \cite{autoran}. Despite this progress, reliably generating correct and standard-compliant configurations remains challenging, especially when tasks involve multi-step workflows, strict syntax or view constraints, and multiple dependent decisions in \cite{netconfeval}. Single-shot or monolithic prompting can lead to error accumulation, generating outputs that may be syntactically plausible but semantically inconsistent or operationally infeasible. More broadly, recent studies suggest that decomposing complex objectives into smaller sub-problems, together with intermediate verification or refinement, can improve reasoning robustness and reduce hallucinations \cite{khot2023decomposed, wei2025inta, dhuliawala2024chain, 10.5555/3666122.3668141}. 

The practical adoption of LLMs in network deployment and management requires more than natural language understanding and structured text generation. Network configuration tasks are highly sensitive to standard compliance, vendor-specific constraints, and platform-dependent implementation. Consequently, the reliability of LLM-generated outputs depends critically on access to accurate and relevant domain knowledge at inference time \cite{LLMNetw}. Retrieval-augmented generation (RAG) naturally extends LLM-assisted network automation by grounding outputs in external knowledge sources such as standards, vendor manuals, operational policies, configuration templates, and validated examples. By retrieving and injecting relevant context before generation, RAG can improve alignment with domain constraints, implementation requirements, and deployment objectives. This is particularly valuable for configuration generation and verification tasks, where correctness depends not only on linguistic coherence but also on syntactic validity, semantic consistency, and platform-specific feasibility \cite{ragnlp, ragtelecom,intentsmorag}.

\begin{figure*}[t]
    \centering
    \includegraphics[width=\textwidth]{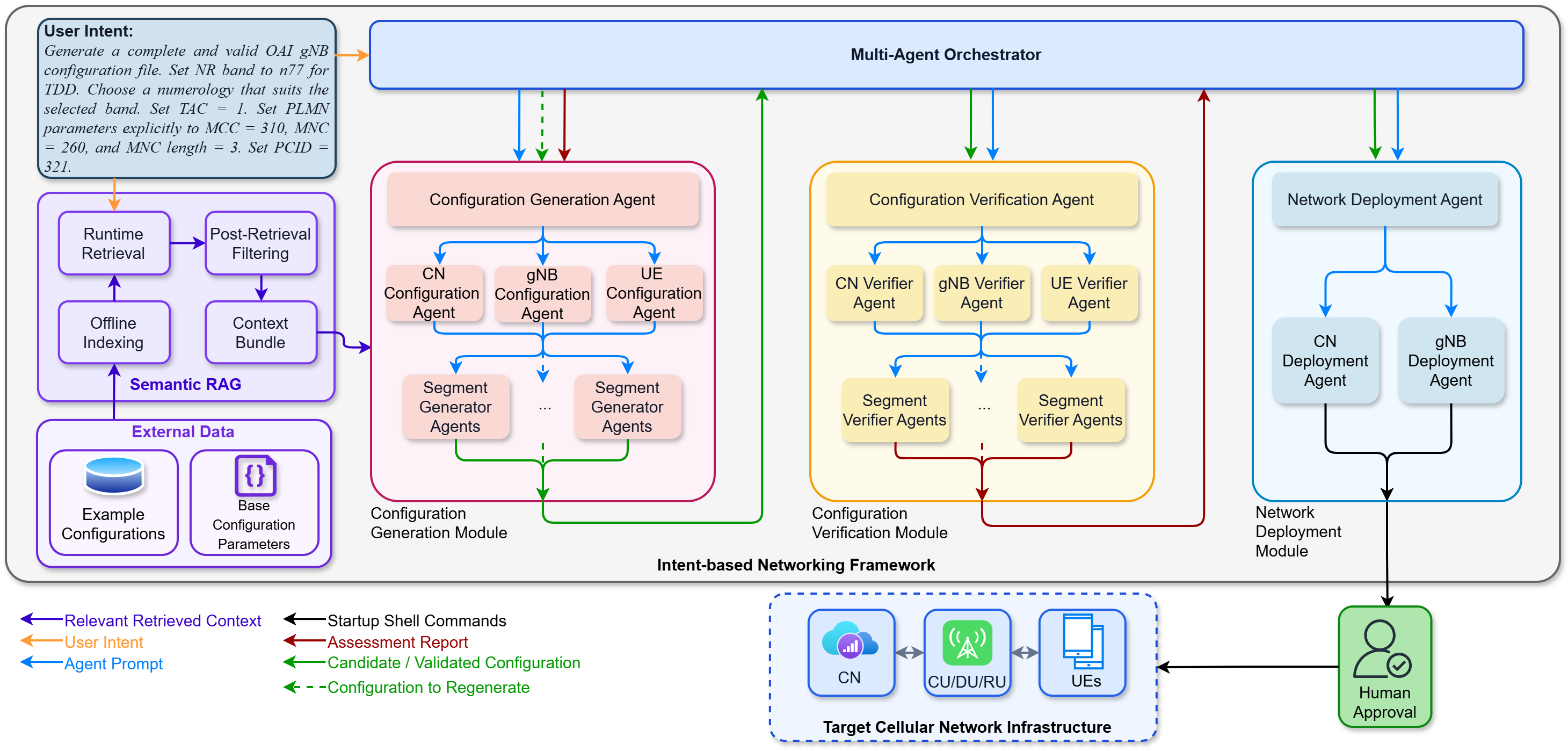}
    \caption{RAG-driven multi-agent framework for intent-based B5G auto-configuration, including configuration generation, verification, and deployment layers.}
    \label{fig:system_model}
\end{figure*}

\needspace{3\baselineskip}

The above-mentioned observations reveal an important gap. Although LLMs show strong potential for intent-driven automation, translating user intent reliably into deployable, standards-aware network configurations still demands both domain grounding and structured reasoning. To address this gap, this paper proposes a retrieval-augmented and decomposed LLM-based framework with human-in-the-loop supervision for generating, verifying, and deploying multi-vendor configurations in B5G cellular networks. Specifically, the proposed approach improves configuration reliability, structural correctness, standard compliance, and deployment readiness by combining RAG, block-structured configuration synthesis, and closed-loop LLM-based verification within an iterative self-refinement workflow, while allowing human experts to review and oversee intermediate outputs when needed. The main contributions of this work can be summarized as follows:

\begin{itemize}

    \item We show that segment-level decomposition combined with selective regeneration is the key mechanism that enables reliable LLM-based network configuration under strict parameter dependencies
    
    \item We introduce a block-structured generation strategy for B5G network configurations, in which complex configuration files are decomposed into logically isolated parameter blocks to reduce hallucination and improve structural correctness during both generation and evaluation. 

    \item We propose a human-in-the-loop verification framework with a closed-loop self-refinement mechanism in which generated configurations are evaluated by a configuration verifier agent powered by an LLM. The system provides structured self-feedback upon failure and iteratively regenerates configurations until compliance criteria are satisfied. This verification pipeline establishes a self-refinement workflow for B5G network configurations, significantly improving reliability, standards compliance, and deployment readiness.

    \item As a further contribution, the proposed RAG-driven agentic LLM framework is implemented in Python and tested using the open source OpenAirInterface (OAI) 5G emulator.
\end{itemize}

The remainder of this paper is structured as follows. Section II introduces the system model. Section III describes the experimental setup and evaluates the performance results. Finally, Section IV concludes the paper.

\section{System Model}
Figure~\ref{fig:system_model} shows the system model of a RAG-driven agentic LLM-based framework for translating natural-language human intents into validated configurations and deployment actions for B5G networks. The framework operates within the service management and orchestration (SMO) platform, where a multi-agent orchestrator coordinates all the operations. First, the framework accepts a natural-language \textit{user intent} as input and transforms it into machine-readable configurations and deployment commands. The \textit{external data} used as a knowledge source contains example configuration files and base configuration parameters. The example files correspond to validated configurations prepared by the vendors or the network operators for diverse deployment settings, including different radio frequency bands, channel bandwidths, numerologies, duplexing schemes, and network slicing profiles, whereas the base configuration parameters define the mandatory parameters that must explicitly appear in the generated outputs. Each parameter includes its formal definition, data type, allowed range, dependency relations, and calculation methodology, such as carrier frequency derivation and raster alignment constraints. This knowledge is processed by the \textit{semantic RAG} pipeline through offline indexing, runtime retrieval, post-retrieval filtering, and context-bundle construction, thereby ensuring that only the most relevant contextual information is supplied to downstream agents.

The \textit{multi-agent orchestrator} supervises the complete workflow by translating the user intent into agent-specific prompts, forwarding the retrieved context to the corresponding agents, collecting the generated configuration files, and processing the assessment reports returned by the verification stage. The multi-agent orchestrator coordinates the activities of three specialized agents, namely the configuration generation agent (CGA), the configuration verification agent (CVA), and the network deployment agent (NDA).  The user intent is forwarded to the CGA, where sub-agents for the CN, gNB, and UE domains generate machine-readable configurations by jointly exploiting the user request and the relevant knowledge retrieved through the RAG pipeline. The generated configurations are subsequently passed to the CVA, where they are validated for structural and semantic correctness using the self-refinement capabilities of LLMs. The CVA generates an assessment report specifying the error status and required revisions and feeds it back to the orchestrator. If any configuration parameters are incorrect, the orchestrator transmits the assessment report and the generated configuration to CGA for reproduction. The workflow then iterates between configuration generation and verification until a valid set of configurations is obtained.

The validated configurations together with the original user intent are forwarded to the NDA, where the CN and gNB deployment agents generate structured startup shell commands for the corresponding network elements according to the user intent. Before deploying on the target infrastructure, all deployment commands are presented to the user for approval as part of the human-in-the-loop control mechanism. Upon approval, the commands are executed, resulting in the deployment and configuration of the target cellular network and enabling subsequent UE attachment and service procedures. Owing to its modular design, CGA, CVA, and NDA can also operate independently, thereby allowing flexible adoption according to the operator's specific requirements while preserving the overall automated yet human-supervised orchestration capability of the framework. The details of each agent are presented below.

\subsection{Configuration Generation}

The configuration generation module is designed as a multi-agent framework which is coordinated by the CGA. It receives the task prompt from the orchestrator agent and enriches the generation prompts with external knowledge through a RAG module to mitigate hallucinations and improve compliance with 3GPP standards. The example configurations are indexed offline and the retriever identifies the most relevant configuration templates at runtime for a given user intent through a top-\(k\) nearest-neighbour search, followed by post-retrieval filtering to remove irrelevant or conflicting items. This retrieved context is then injected into the prompts to keep the generated outputs consistent with the valid parameter ranges and dependency constraints. In this module, three specialized configuration agents are defined for the gNB, CN, and UE domains. Each specialized agent decomposes its assigned domain-specific task into logical parameter groups and assigns them to segment generator agents for concurrent generation. For example, within the gNB domain, the assigned task may be divided into CU-, DU-, and RU-related parameter groups, each of which is handled by a separate segment generation agent. More generally, the decomposition follows the functional structure of the target domain, such as CU, DU, and RU parameter groups in the gNB, or the parameter sets associated with individual CN functions. Each segment generation agent generates a strictly formatted configuration segment for its assigned partition. Once all segments are generated, the corresponding specialized agent aggregates them into a domain-level configuration. Finally, the coordinator combines the outputs of the CN, gNB, and UE branches into a candidate configuration package for the verification and deployment stages. Note that, in a monolithic approach, each specialized agent generates the configuration of its assigned domain in a single pass, without further partitioning the task into smaller sub-tasks.

\subsection{Configuration Verification}
The configuration verification module is responsible for validating the generated configurations and identifying structural and semantic inconsistencies before deployment. In this module, the CVA examines the generated configurations for missing mandatory parameters, invalid or out-of-range values, inter-parameter dependency violations, and syntax errors. In the decomposition-based verification approach, the network configuration is partitioned into logically coherent segments corresponding to independent functional blocks, following the same decomposition logic adopted in the configuration generation module. This partitioning is performed by a Python function invoked by each specialized verifier agent, generating smaller and semantically consistent verification units that are assigned to segment verifier agents. They validate their assigned segments concurrently and generate structured assessments, each consisting of a binary validity indicator and a list of detected issues. The verifier agent aggregates these outputs, resolves possible inter-segment dependency conflicts, and produces a consolidated assessment report in JSON format. For valid segments, the report explicitly states that no correction is required. For invalid segments, it identifies the faulty parameter, explains the root cause of the issue, and provides a suggested correction strategy. 

Note that, in the monolithic approach, the CVA receives the complete configuration as a single input and performs the validation in one pass. In this case, even a single detected error often necessitates regenerating the entire configuration, resulting in unnecessary computational overhead and higher token consumption. A key advantage of the decomposition-based verification approach is that only the faulty segments identified during verification are sent back for regeneration, while the valid segments are preserved and reused.  By enabling selective regeneration, the proposed mechanism improves the efficiency of the closed-loop configuration workflow while reducing both computational cost and token usage. The assessment report is then sent back to the configuration generation module, where the corresponding configuration agents regenerate only the affected segments by incorporating the provided feedback. The updated configuration is subsequently resubmitted to the configuration verification module, and this iterative process continues until all structural and semantic constraints are satisfied. Once no issues remain, the configuration is deemed valid, stored for deployment, and forwarded to the orchestrator agent for the subsequent deployment stage.

\subsection{Network Deployment}

In the network deployment module, executable commands that include the validated configurations for the target network infrastructure are generated. The NDA manages two specialized agents, where the gNB deployment agent generates the deployment plan required to instantiate the CU, DU, and RU components together with their associated configuration files, whereas the CN deployment agent generates the deployment plan for the CN functions. Both agents take the user intent and the references to the validated configuration as inputs and generate structured shell commands for their respective network elements. To avoid the autonomous execution of unsafe or unintended actions, these commands are not executed directly. Instead, they are presented to the user for approval before execution. This human-in-the-loop mechanism improves the reliability and safety of the deployment process by ensuring that critical deployment actions remain subject to human oversight. Once the generated shell commands are approved, the corresponding deployment agent accesses the terminal interface of the target network element through the Python-based control functions and executes the generated shell commands in the target environment.

\section{Implementation and Performance Evaluation}

\subsection{Experimental Setup}
The proposed system is implemented over a private 5G network built on the OAI \cite{10.1145/2677046.2677053}, which provides software implementations of the 5GC, gNB, and UE. The OAI radio frequency (RF) simulator is employed for setting up an end-to-end network emulation. The system is deployed on Supermicro ARS-111GL-NHR server running on Ubuntu Server 22.04, which also supports local LLM inference. In this environment, the GPT-OSS-120B model is executed through Ollama platform and integrated with the Python-based orchestration framework via LangChain. On the retrieval side, FAISS is utilized for vector similarity search, while all-MiniLM-L6-v2 which is an embedding model in Hugging Face is used mapping text to vector space. The RAG knowledge base consists of 37 gNB and 5 UE configuration files obtained from the official OAI repository. The temperature parameter is set to 0.3 to balance between deterministic structured output generation with sufficient probabilistic flexibility, allowing the model to infer appropriate parameter values from the user intent while preserving structural consistency. The implementation environment and software stack are summarized in Table \ref{tab:experimental_platform}. For the sake of demonstration, the framework is utilized to automatically generate, verify, and deploy the gNB configurations while the default CN and UE configurations are used for all experiments. 

\subsection{Experimental Results}

\begin{table}[t]
\caption{Experimental platform and software stack.}
\label{tab:experimental_platform}
\centering
\begin{tabular}{|>{\centering\arraybackslash}m{4cm}|>{\centering\arraybackslash}m{4cm}|}
\hline
\textbf{Item} & \textbf{Description} \\
\hline
5G network & OpenAirInterface \\
\hline
Operating system & Ubuntu Server 22.04 \\
\hline
CPU & 72-core ARM CPU \\
\hline
GPU & NVIDIA H100 \\
\hline
LLM platform & Ollama \\
\hline
LLM model & GPT-OSS-120B \\
\hline
Orchestration & LangChain \\
\hline
Vector similarity engine & FAISS \\
\hline
Embedding models & all-MiniLM-L6-v2 \\
\hline
Temperature & 0.3 \\
\hline
\end{tabular}
\end{table}

The performance evaluation assesses the correctness of the generated configurations both without and with verification, while comparing the monolithic and decomposed approaches. For each approach, 10 different intent prompts are defined, and 100 configurations are generated per prompt, resulting in 1000 configurations per strategy and 2000 generated gNB configurations in total. To ensure a fair comparison, the same prompt set is used for each method. Although prompts are synthetically generated with LLM assistance to increase diversity, they are derived from real configuration parameters pool and still reflect valid operational constraints. For each prompt, the LLM randomly selects five parameters from the pool and assigns valid random values to construct a natural-language configuration request. An example prompt used for configuration generation is as follows:

\vspace{0.6em}
\textit{
"Generate a complete and valid OAI gNB configuration file.
Set NR band to n77 for TDD.
Choose a numerology that suits the selected band.
Set TAC = 1.
Set PLMN parameters explicitly to MCC = 310, MNC = 260, and MNC length = 3.
Set PCID = 321."
}
\vspace{0.6em}

\begin{figure}
    \centering
    \includegraphics[width=\linewidth]{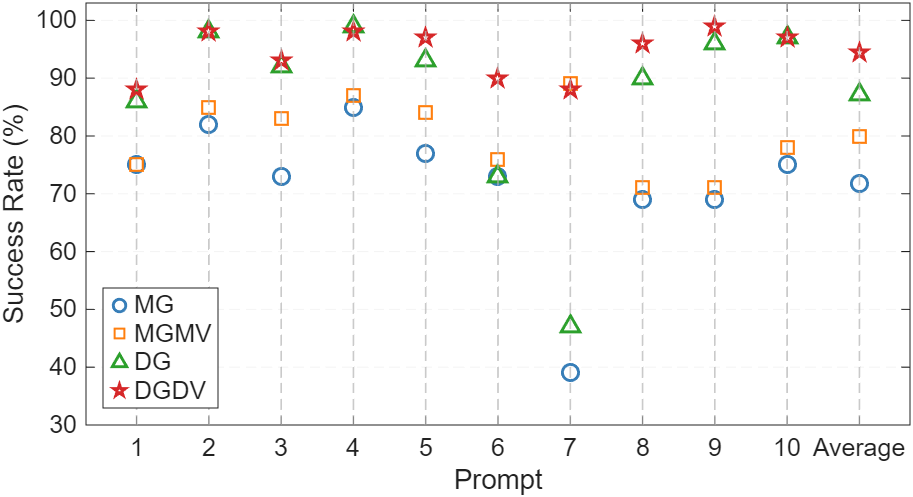}
    \caption{Performance comparison of MG, MGMV, DG, and DGDV across different prompts.}
    \label{fig:sonuc1}
\end{figure}

Each of the 2,000 generated configurations is then deployed in the RF simulator environment using the OAI gNB stack, under both monolithic and decomposed approaches. The deployment success rate for each approach is computed as
\begin{equation}
    {\textit{Success Rate}}= \frac{N_{\mathrm{success}}}{N_{\mathrm{total}}} \times 100,
    \label{eq:successrate}
\end{equation}
where $N_{\mathrm{success}}$ and $N_{\mathrm{total}}$ denote the number of successfully deployed configurations and the total number of generated configurations, respectively. A configuration is regarded as successful only if it is aligned with the user intent and results in an error-free gNB deployment.

Figure \ref{fig:sonuc1} shows the generation success rate results across four methods, namely monolithic generation (MG), decomposed generation (DG), monolithic generation monolithic verification (MGMV), and decomposed generation decomposed verification (DGDV). MG represents single-pass configuration generation using a RAG-enhanced LLM, while DG denotes configuration generation performed through task decomposition into smaller sub-tasks. MGMV corresponds to monolithic generation followed by monolithic verification, where the entire configuration is validated in a single step. DGDV represents decomposed generation combined with decomposed verification, where both generation and validation are carried out at the segment level. The figure illustrates the success rates of 10 different prompts, with the last one representing the average success rate, and presents an ablation study comparing the MG–DG, MG–MGMV, and MG–DGDV approaches. The results indicate that the DG approach consistently outperforms the MG approach, achieving a 15.4\% improvement in the average success rate. This suggests that decomposing the complex configuration task into smaller, domain-specific sub-problems can improve the effectiveness of the LLM-based framework in strict network configuration synthesis. However, this improvement comes at the cost of higher latency. 

As reported in Table \ref{tab:inference_all}, the mean inference time defined as the time required for an LLM to generate outputs from input prompts, increases from 18.274~seconds for MG to 47.055~seconds for the DG, reflecting the overhead of coordinating multiple agent calls. The effect of the verification stage after the proposed generation approach is also illustrated in Fig. \ref{fig:sonuc1}. To assess the impact of self-refinement independent from the generation strategy, configurations generated with MG are first subjected to a MV process. The results show that verification improves the average success rate of MG by 8.2\%, demonstrating that the self-refinement capability of LLMs can effectively identify and correct errors introduced during the initial generation stage. Finally, DGDV yields the highest average success rate of 94.4\%, confirming that task decomposition and self-refinement are complementary mechanisms that together substantially enhance the reliability of the overall framework. As shown in Fig.~\ref{fig:sonuc1}, both the MG and DG exhibit relatively lower success rates for Prompt 7 compared to the other prompts. This degradation is attributed to conflicts between the system prompt defined via prompt engineering, the retrieved contextual information, and the knowledge embedded in the pre-trained LLM. Such inconsistencies may have contributed to hallucinations during generation, particularly for interdependent configuration parameters, leading to invalid outputs. In this case, the role of the CVA, which utilizes the self-refinement capability of LLMs, becomes more apparent. The MV increases the success rate of MG from 39\% to 89\%, while the DV improves the success rate of DG from 47\% to 88\%. These observations suggest that the verification stage helps recover a substantial portion of the initially unsuccessful configurations and preserves the overall success trend of the verified outputs.

The latency trade-off introduced by the verification stage is also summarized in Table \ref{tab:inference_all}. The results show that MV requires 27.612~seconds on average, whereas DV requires 32.905~seconds. Thus, DV introduces additional latency, although this overhead remains limited relative to the reliability gains provided by the overall decomposed workflow. When both generation and verification times are considered together, the mean total inference time increases from 45.886~seconds for the monolithic pipeline to 79.961~seconds for the decomposed pipeline. Therefore, the proposed decomposed framework achieves higher deployment success at the expense of a longer end-to-end inference time.

\begin{table}[t]
\caption{Inference time statistics for monolithic and decomposition-based approaches.}
\label{tab:inference_all}
\centering
\footnotesize
\begin{tabular}{|>{\centering\arraybackslash}m{1.4cm}|>{\centering\arraybackslash}m{1.2cm}|>{\centering\arraybackslash}m{1.6cm}|>{\centering\arraybackslash}m{1.6cm}|>{\centering\arraybackslash}m{0.9cm}|}
\hline
\textbf{Approach} & \textbf{Statistic (Seconds)} & \textbf{Configuration Generation} & \textbf{Configuration Verification} & \textbf{Overall} \\
\hline
\multirow{2}{*}{\textbf{Monolithic}}
& Mean & 18.274 & 27.612 & 45.886 \\
\cline{2-5}
& Std. & 3.582 & 4.824 & 5.918 \\
\hline
\multirow{2}{*}{\textbf{Decomposed}}
& Mean & 47.055 & 32.905 & 79.961 \\
\cline{2-5}
& Std. & 4.964 & 3.527 & 5.668 \\
\hline
\end{tabular}
\end{table}

\begin{table}[h!]
\caption{Error distributions of the configuration generation approaches.}
\label{tab:error_distribution}
\centering
\begin{tabular}{|>{\centering\arraybackslash}m{1.4cm}|>{\centering\arraybackslash}m{1.1cm}|>{\centering\arraybackslash}m{1cm}|>{\centering\arraybackslash}m{1.6cm}|>{\centering\arraybackslash}m{1.6cm}|}
\hline
\textbf{Error Type} & \textbf{MG (\%)} & \textbf{DG (\%)} & \textbf{MGMV (\%)} & \textbf{DGDV (\%)} \\
\hline 
Syntax error         & 7.66  & 29.73 & 17.86 & 0.55 \\ 
\hline
Duplicated parameter & 0.45  & 35.14 & 0.00  & 0.00 \\
\hline
Invalid values    & 91.89  & 35.14 & 82.14 & 99.45 \\
\hline
\end{tabular}
\end{table}

The distribution of errors summarized in Table \ref{tab:error_distribution} indicates that the failures are mainly caused by invalid values across most methods, while duplicated parameters and syntax errors contribute notably in DG. This indicates that the hallucination-related limitations remain in domain-constrained generation tasks, despite the benefits of prompt engineering and RAG. Prompt engineering offers time and resource savings by avoiding model retraining, but this efficiency comes with limited domain fidelity under strict parameter dependencies and implementation constraints. 

Overall, the DGDV approach is more reliable for strictly formatted configuration generation because it alleviates the burden on the LLM of generating a fully valid long-form configuration in a single pass. Furthermore, the verifier stage acts as a self-refinement mechanism that enhances both syntactic and semantic consistency prior to deployment. This improved reliability, however, comes at the expense of higher inference latency, as DGDV requires multiple coordinated model calls. 

\section{Conclusion}
This paper presents a retrieval-augmented and task decomposition-based multi-agent LLM framework for B5G network auto-configuration. The proposed system introduces a layered architecture consisting of configuration generation, closed-loop self-refinement-based verification, and network deployment, enabling the transformation of human intent into machine-readable configurations and executable deployment commands. By leveraging prompt engineering and retrieval augmentation, the framework reduces development effort and operational overhead. The system is validated using the OAI 5G emulator, demonstrating that, despite an increase in inference time, the proposed decomposition-based configuration and verification approaches improve the average success rate by 22.7\% over monolithic methods, achieving 94.4\% success in network configuration. 

This study reveals several valuable opportunities for further research. First, the proposed framework leverages prompt engineering and structured task decomposition. While the current decomposition is static, it provides a strong foundation that can be extended toward greater portability across diverse networks, including different vendors, emulators, and deployment environments. Future work can therefore explore autonomous task decomposition by the LLM, where configuration parameters are grouped according to their semantic and functional dependencies rather than fixed line ranges. Second, the current evaluation is limited to GPT-OSS-120B. A broader benchmark involving multiple LLMs is needed to identify the most suitable model for each agent and to support an AI-as-a-Service (AIaaS) deployment paradigm. Finally, explainable LLM mechanisms should be integrated to improve transparency and reduce false alarms and misdetections.

\bibliographystyle{IEEEtran}
\bibliography{main.bib}

@IEEEtranBSTCTL{IEEEexample:BSTcontrol,
CTLuse_forced_etal       = "yes",
CTLmax_names_forced_etal = "8",
CTLnames_show_etal       = "1" }

@inproceedings{
khot2023decomposed,
title={{D}ecomposed {P}rompting: {A} {M}odular {A}pproach for {S}olving {C}omplex {T}asks},
author={Tushar Khot and Harsh Trivedi and Matthew Finlayson and Yao Fu and Kyle Richardson and Peter Clark and Ashish Sabharwal},
booktitle={{T}he {E}leventh {I}nternational {C}onference on {L}earning {R}epresentations},
year={2023},
}

@ARTICLE{anm2012,
  author={Movahedi, Zeinab and Ayari, Mouna and Langar, Rami and Pujolle, Guy},
  journal={{IEEE} {C}ommunications {S}urveys \& {T}utorials},
  title={{A} {S}urvey of {A}utonomic {N}etwork {A}rchitectures and {E}valuation {C}riteria},
  year={2012},
  volume={14},
  number={2},
  pages={464-490},
  doi={10.1109/SURV.2011.042711.00078}}

@article{etsi2019zero,
  title={{Z}ero-{T}ouch {N}etwork and {S}ervice {M}anagement ({ZSM}): {R}eference {A}rchitecture},
  author={ETSI GS ZSM},
  journal={{ETSI} {G}roup {S}pecification},
  volume={2},
  year={2019}
}

@ARTICLE{llmnetworking,
  author={Huang, Yudong and Du, Hongyang and Zhang, Xinyuan and Niyato, Dusit and Kang, Jiawen and Xiong, Zehui and Wang, Shuo and Huang, Tao},
  journal={{IEEE} {N}etwork},
  title={{L}arge {L}anguage {M}odels for {N}etworking: {A}pplications, {E}nabling {T}echniques, and {C}hallenges},
  year={2025},
  volume={39},
  number={1},
  pages={235-242},
  doi={10.1109/MNET.2024.3435752}}

@article{mekrache2024intent,
  title={{I}ntent-{B}ased {M}anagement of {N}ext-{G}eneration {N}etworks: {A}n {LLM}-{C}entric {A}pproach},
  author={Mekrache, Abdelkader and Ksentini, Adlen and Verikoukis, Christos},
  journal={{IEEE} {N}etwork},
  volume={38},
  number={5},
  pages={29--36},
  year={2024},
  publisher={{IEEE}}
}

@article{coronado2022zero,
  title={{Z}ero-{T}ouch {M}anagement: {A} {S}urvey of {N}etwork {A}utomation {S}olutions for {5G} and {6G} {N}etworks},
  author={Coronado, Estefania and Behravesh, Rasoul and Subramanya, Tejas and Fernandez-Fernandez, Adriana and Siddiqui, Muhammad Shuaib and Costa-P{\'e}rez, Xavier and Riggio, Roberto},
  journal={{IEEE} {C}ommunications {S}urveys \& {T}utorials},
  volume={24},
  number={4},
  pages={2535--2578},
  year={2022},
  publisher={{IEEE}}
}

@ARTICLE{LLMNetConf,
  author={Lira, Oscar G. and Caicedo, Oscar M. and da Fonseca, Nelson L. S.},
  journal={{IEEE} {C}ommunications {M}agazine},
  title={{L}arge {L}anguage {M}odels for {Z}ero {T}ouch {N}etwork {C}onfiguration {M}anagement},
  year={2025},
  volume={63},
  number={7},
  pages={146-153},
  doi={10.1109/MCOM.001.2400368}}

@article{LLMNetw,
  title={{L}arge {L}anguage {M}odels for {N}etworking: {A}pplications, {E}nabling {T}echniques, and {C}hallenges},
  author={Huang, Yudong and Du, Hongyang and Zhang, Xinyuan and Niyato, Dusit and Kang, Jiawen and Xiong, Zehui and Wang, Shuo and Huang, Tao},
  journal={{IEEE} {N}etwork},
  volume={39},
  number={1},
  pages={235--242},
  year={2024},
  publisher={{IEEE}}
}

@article{ragtelecom,
  title={{T}elecom{RAG}: {T}aming {T}elecom {S}tandards with {R}etrieval-{A}ugmented {G}eneration and {LLM}s},
  author={Yilma, Girma M and Ayala-Romero, Jose A and Garcia-Saavedra, Andres and Costa-Perez, Xavier},
  journal={{ACM} {SIGCOMM} {C}omputer {C}ommunication {R}eview},
  volume={54},
  number={3},
  pages={18--23},
  year={2025},
  publisher={{ACM}}
}

@misc{agentran,
      title={{A}gent{RAN}: {A}n {A}gentic {AI} {A}rchitecture for {A}utonomous {C}ontrol of {O}pen {6G} {N}etworks},
      author={Maxime Elkael and Salvatore D'Oro and Leonardo Bonati and Michele Polese and Yunseong Lee and Koichiro Furueda and Tommaso Melodia},
      year={2026},
      eprint={2508.17778},
      archivePrefix={{arXiv}},
      primaryClass={{cs.AI}},
      url={https://arxiv.org/abs/2508.17778},
}

@ARTICLE{autoran,
  author={Maxenti, Stefano and Shirkhani, Ravis and Elkael, Maxime and Bonati, Leonardo and D'Oro, Salvatore and Melodia, Tommaso and Polese, Michele},
  journal={{IEEE} {T}ransactions on {M}obile {C}omputing},
  title={{A}uto{RAN}: {A}utomated and {Z}ero-{T}ouch {O}pen {RAN} {S}ystems},
  year={2026},
  volume={},
  number={},
  pages={1-18},
  doi={10.1109/TMC.2026.3656056}
  }

@article{netconfeval,
  title={{N}et{C}onf{E}val: {C}an {LLM}s {F}acilitate {N}etwork {C}onfiguration?},
  author={Wang, Changjie and Scazzariello, Mariano and Farshin, Alireza and Ferlin, Simone and Kosti{\'c}, Dejan and Chiesa, Marco},
  journal={{P}roceedings of the {ACM} on {N}etworking},
  volume={2},
  number={{CoNEXT2}},
  pages={1--25},
  year={2024},
  publisher={{ACM}}
}

@inproceedings{wei2025inta,
  title={{INTA}: {I}ntent-{B}ased {T}ranslation for {N}etwork {C}onfiguration with {LLM} {A}gents},
  author={Wei, Yunze and Xie, Xiaohui and Hu, Tianshuo and Zuo, Yiwei and Chen, Xinyi and Chi, Kaiwen and Cui, Yong},
  booktitle={{2025} {IEEE} {33rd} {I}nternational {C}onference on {N}etwork {P}rotocols ({ICNP})},
  pages={1--16},
  year={2025},
  organization={{IEEE}}
}

@inproceedings{dhuliawala2024chain,
  title={{C}hain-of-{V}erification {R}educes {H}allucination in {L}arge {L}anguage {M}odels},
  author={Dhuliawala, Shehzaad and Komeili, Mojtaba and Xu, Jing and Raileanu, Roberta and Li, Xian and Celikyilmaz, Asli and Weston, Jason},
  booktitle={{F}indings of the {A}ssociation for {C}omputational {L}inguistics: {ACL} {2024}},
  pages={3563--3578},
  year={2024}
}

@inproceedings{intentsmorag,
  title={{I}ntent {B}ased {N}etworking for {S}ervice {M}anagement \& {O}rchestration of {5G} {N}etworks},
  author={Rushiti, Veton and Jayakumar, Bharani and Shaik, Zubair and Mitschele-Thiel, Andreas and Parameswaran, Sriram},
  booktitle={{2025} {IEEE} {36th} {I}nternational {S}ymposium on {P}ersonal, {I}ndoor and {M}obile {R}adio {C}ommunications ({PIMRC})},
  pages={1--6},
  year={2025},
  organization={{IEEE}}
}

@misc{ITU2026IMT2030,
  author       = {{International Telecommunication Union}},
  title        = {{IMT-2030}: {T}echnical {R}equirements for the {6G} {F}uture},
  year         = {2026},
  month        = mar,
  note         = {{ITU} {N}ews, accessed: 2026-03-23},
  url          = {https://www.itu.int/hub/2026/03/imt-2030-technical-requirements-for-the-6g-future/}
}

@inproceedings{ragnlp,
author = {Lewis, Patrick and Perez, Ethan and Piktus, Aleksandra and Petroni, Fabio and Karpukhin, Vladimir and Goyal, Naman and K\"{u}ttler, Heinrich and Lewis, Mike and Yih, Wen-tau and Rockt\"{a}schel, Tim and Riedel, Sebastian and Kiela, Douwe},
title={{R}etrieval-{A}ugmented {G}eneration for {K}nowledge-{I}ntensive {NLP} {T}asks},
year = {2020},
isbn = {9781713829546},
publisher = {{C}urran {A}ssociates {I}nc.},
address = {{R}ed {H}ook, {NY}, {USA}},
booktitle={{P}roceedings of the {34th} {I}nternational {C}onference on {N}eural {I}nformation {P}rocessing {S}ystems},
articleno = {793},
numpages = {16},
location = {{V}ancouver, {BC}, {C}anada},
series = {{NIPS} '20}
}

@ARTICLE{llmfortelecom,
  author={Zhou, Hao and Hu, Chengming and Yuan, Ye and Cui, Yufei and Jin, Yili and Chen, Can and Wu, Haolun and Yuan, Dun and Jiang, Li and Wu, Di and Liu, Xue and Zhang, Jianzhong and Wang, Xianbin and Liu, Jiangchuan},
  journal={{IEEE} {C}ommunications {S}urveys \& {T}utorials},
  title={{L}arge {L}anguage {M}odel ({LLM}) for {T}elecommunications: {A} {C}omprehensive {S}urvey on {P}rinciples, {K}ey {T}echniques, and {O}pportunities},
  year={2025},
  volume={27},
  number={3},
  pages={1955-2005},
  doi={10.1109/COMST.2024.3465447}}

@inproceedings{10.5555/3666122.3668141,
author = {Madaan, Aman and Tandon, Niket and Gupta, Prakhar and Hallinan, Skyler and Gao, Luyu and Wiegreffe, Sarah and Alon, Uri and Dziri, Nouha and Prabhumoye, Shrimai and Yang, Yiming and Gupta, Shashank and Majumder, Bodhisattwa Prasad and Hermann, Katherine and Welleck, Sean and Yazdanbakhsh, Amir and Clark, Peter},
title = {{SELF}-{REFINE}: {I}terative {R}efinement with {S}elf-{F}eedback},
year = {2023},
publisher = {{C}urran {A}ssociates {I}nc.},
address = {{R}ed {H}ook, {NY}, {USA}},
booktitle = {{P}roceedings of the {37th} {I}nternational {C}onference on {N}eural {I}nformation {P}rocessing {S}ystems},
articleno = {2019},
numpages = {61},
location = {{N}ew {O}rleans, {LA}, {USA}},
series = {{NIPS} '23}
}

@article{10.1145/2677046.2677053,
author = {Nikaein, Navid and Marina, Mahesh K. and Manickam, Saravana and Dawson, Alex and Knopp, Raymond and Bonnet, Christian},
title = {{O}pen{A}ir{I}nterface: {A} {F}lexible {P}latform for {5G} {R}esearch},
year = {2014},
issue_date = {October 2014},
publisher = {{A}ssociation for {C}omputing {M}achinery},
address = {{N}ew {Y}ork, {NY}, {USA}},
volume = {44},
number = {5},
issn = {0146-4833},
url = {https://doi.org/10.1145/2677046.2677053},
doi = {10.1145/2677046.2677053},
journal = {{SIGCOMM} {C}omput. {C}ommun. {R}ev.},
month = oct,
pages = {33–38},
numpages = {6}
}
\end{document}